\documentclass[twocolumn,showpacs,prl,amsmath,amssymb]{revtex4}
\usepackage{graphicx}
\usepackage{dcolumn}
\usepackage{bm}
\usepackage{textcomp}

\begin{document}
\title{Non-adiabatic Kohn-anomaly in a doped graphene monolayer}

\author{Michele Lazzeri and Francesco Mauri}
\affiliation{
IMPMC, Universit\'es Paris 6 et 7, CNRS, IPGP,
140 rue de Lourmel, 75015 Paris, France}
\date{\today}

\begin{abstract}
We compute, from first-principles, the frequency of the E$_{2g}$,
${\bm \Gamma}$ phonon (Raman $G$ band) of graphene, as a function of
the charge doping.  Calculations are done using i) the {\it adiabatic}
Born-Oppenheimer approximation and ii) time-dependent perturbation theory
to explore {\it dynamic} effects beyond this approximation. The two
approaches provide very different results.  While, the {\it adiabatic}
phonon frequency weakly depends on the doping, the {\it dynamic} one
rapidly varies because of a Kohn anomaly.  The {\it adiabatic}
approximation is considered valid in most materials.  Here, we show
that doped graphene is a spectacular example where this approximation
miserably fails.
\end{abstract}
                 
\maketitle

Graphene is a 2-dimensional plane of carbon atoms arranged in a
honeycomb lattice. The recent demonstration of a field-effect
transistor (FET) based on a few-layers graphene sheet has boosted the
interest in this system~\cite{geim,kim05,ferrari06}.  In particular,
by tuning the FET gate-voltage $V_g$ it is possible to dope graphene by
adding an excess surface electron charge.  The actual possibility of
building a FET with just one graphene monolayer maximizes the excess
charge corresponding to a single atom in the sheet.  In a FET-based
experiment, graphene can be doped up to $3~10^{13}$ ~cm$^{-2}$
electron concentration~\cite{geim,kim05}, corresponding, in a
monolayer, to a 0.2\% valence charge variation.
The resulting chemical-bond modification could induce a
variation of bond-lengths and phonon-frequencies of the same
order, which would be measurable.  This would
realize the dream of tuning the chemistry, within an electronic
device, by varying $V_g$.

The presence of Kohn anomalies (KAs)~\cite{kohn59,piscanec04}
in graphene could act as a magnifying glass, leading to a variation
of the optical phonon-frequencies much larger than the 0.2\% expected
in conventional systems.
On the other hand, the phonon-frequency change induced by FET-doping
could provide a much more precise determination of the KA, with 
respect to other experimental settings.
KAs manifest as a sudden change in the phonon dispersion
for a wavevector ${\bf q}\sim2{\bf k}_F$, where ${\bf k}_F$ is
a Fermi-surface wavevector~\cite{kohn59}.  The KA can be
determined by studying the phonon frequency as a function of ${\bf q}$
by, e.g., inelastic x-ray, or neutron scattering.
These techniques
have a finite resolution, in ${\bf q}$ and energy, which limits the
precision on the measured KA dispersion.
In graphene, $2{\bf k}_F$ is proportional to $V_g$. This suggests an
alternative way to study the KA, that is to measure the phonon 
frequency at a fixed ${\bf q}$ and to vary $2{\bf k}_F$ by changing $V_g$.
Within this approach, one could use Raman scattering, which has a much
better energy and momentum resolution than x-ray and neutron scattering.
This approach is feasible for graphene, which has a KA for the
Raman-active E$_{2g}$ ${\bm \Gamma}$-phonon~\cite{piscanec04} (Raman $G$-band).

In this paper, we compute the variation of phonon frequency of the
Raman $G$-band (E$_{2g}$ mode at ${\bm \Gamma}$) in a graphene
monolayer, as a function of the Fermi level.  First, the calculations
are done using a fully ab-initio approach within the customary
adiabatic Born-Oppenheimer approximation.  Then, time-dependent
perturbation theory (TDPT) is used to go beyond.

Ab-initio calculations are done within density functional theory
(DFT), using the functional of Ref.~\cite{pbe}, plane waves (30~Ry cutoff)
and pseudopotentials~\cite{vanderbilt}. 
The Brillouin zone (BZ) integration is done on a uniform $64\times64\times1$
grid.  An electronic smearing of
0.01~Ry with the Fermi-Dirac distribution is
used~\cite{methfessel}.  The two-dimensional graphene crystal is
simulated using a super-cell geometry with an interlayer spacing of 7.5~\AA~
(if not otherwise stated).  Phonon frequencies are
calculated within the approach of Ref.~\cite{dfpt}, using the PWSCF
code~\cite{pwscf}.  The Fermi-energy shift is simulated by
considering an excess electronic charge which is compensated by a
uniformly charged back-ground.

\begin{figure} [h]
\centerline{\includegraphics[width=65mm]{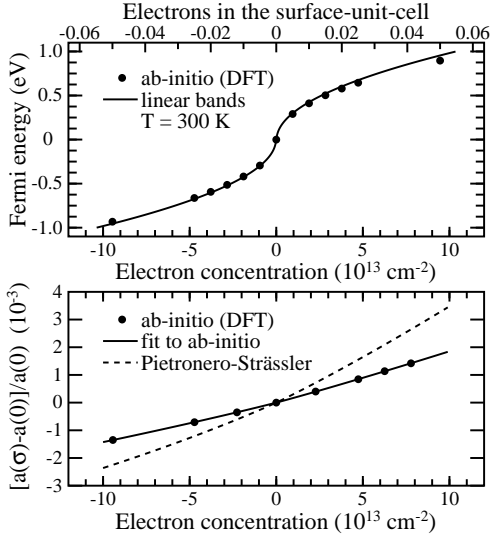}}
\caption{
Graphene monolayer.
Upper panel: $\epsilon_F$ as a function of the surface electron-concentration $\sigma$
from DFT calculations and from linearized bands (at $T=300$~K).
Lower panel: in-plane lattice spacing $a$ as a function of $\sigma$.
The fitting function is Eq.~\protect\ref{eq2}
and the dashed line is from Ref.~\protect\cite{pietronero81}.
}
\label{fig1}
\end{figure}

The dependence of the Fermi energy $\epsilon_F$ on the surface electron-
concentration $\sigma$ is determined by DFT (Fig~\ref{fig1}).
In graphene, the gap is zero only for the two equivalent {\bf K} and
{\bf K'}~BZ-points and the electron energy $\epsilon$
can be approximated as
$\epsilon_{\pi^*/\pi}({\rm \bf K}+{\bf k})=\pm\beta k$
for the $\pi^*$ and $\pi$ bands, where ${\bf k}$ is a small vector.
Within this approximation, at $T$ = 0 K temperature
\begin{equation}
\sigma = {\rm sign}(\epsilon_F) 
\frac{\epsilon_F^2}{\pi\beta^2} =
{\rm sign}(\epsilon_F)
\frac{\epsilon_F^2}{{\rm eV}^2}
~10.36~10^{13}{\rm cm}^{-2}
\label{eq1}
\end{equation}
where $\beta=5.52$~eV$\cdot$\AA~from DFT, ${\rm sign}(x)$ is the sign of $x$
and $\epsilon_F=0$ at the $\pi$ bands crossing.
We remark that, from Fig~\ref{fig1}, the typical
electron-concentration obtained in experiments~\cite{geim,kim05} corresponds to an important
Fermi-level shift ($\sim$0.5~eV). 
For such shift, the linearized bands are still a good approximation (Fig.~\ref{fig1}).


The dependence of the graphene lattice-spacing $a$ on $\sigma$, $a(\sigma)$, is
obtained by minimizing
$F(\sigma,A)=[E(\sigma,A)-E(0,A_0)]/A$
with respect to $A$, where $E(\sigma,A)$ is the energy
of the graphene unit-cell, $A$ is unit-cell area and $A_0=5.29$~\AA$^2$
is the equilibrium $A$~\cite{nota1} at zero $\sigma$.
$E(\sigma,A)$ is computed by DFT letting
the inter-layer spacing, $L$, tend to infinity in order to eliminate the
spurious interaction between the background and the charged sheet~\cite{nota2}.
$\Delta a(\sigma)=[a(\sigma)-a(0)]/a(0)$ was determined in Ref.~\cite{pietronero81}
for intecalated graphite on the basis of a semi empirical model.
Using the same functional dependence
as in Ref.~\cite{pietronero81}, our DFT calculations are
fitted by
\begin{equation}
\Delta a(\sigma) = 6.748 \cdot 10^{-6} |\sigma|^{3/2} + 1.64  \cdot 10^{-4} \sigma,
\label{eq2}
\end{equation}
where $\sigma$ is in units of 10$^{13}$~cm$^{-2}$.
With $\sigma = 3~10^{13}$~cm$^{-2}$, the lattice spacing variation is $\sim 0.05 \%$,
which is, as expected, of the same order of the valence-charge variation.

\begin{figure} [h]
\centerline{\includegraphics[width=75mm]{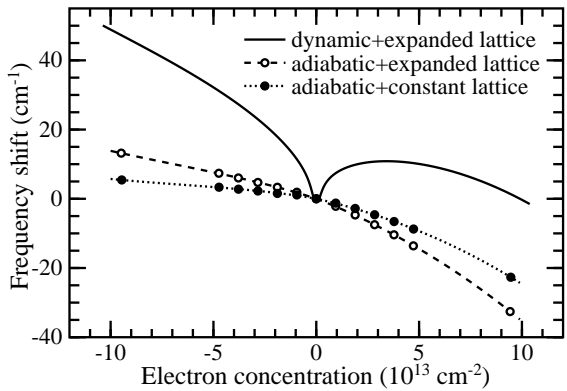}}
\caption{Frequency of the E$_{2g}$ ${\bm \Gamma}$ phonon
(Raman $G$ band) as a function of $\sigma$:
shift with respect to the zero-doping frequency.
Calculations are done using standard DFT (adiabatic)
or TDPT (dynamic), keeping the lattice-spacing constant
(constant lattice) or varying it according to Eq.~\protect\ref{eq2} (expanded lattice).
Points are DFT calculations. Dashed line is from Eq.~\ref{eq2_5}.
Experiments should be compared with the continuous line.
}
\label{fig2}
\end{figure}

The frequency of the E$_{2g}$ ${\bm \Gamma}$ phonon is computed
by {\it static} perturbation theory of the DFT energy~\cite{dfpt},
i.e. from the linearized
forces acting on the atoms due to the {\it static} displacement of
the other atoms from their equilibrium positions.
This approach is based on the {\it adiabatic} Born-Oppenheimer approximation, 
which is the standard
textbook approach for phonon calculations and is always used,
to our knowledge, in the ab-initio frequency calculations.
The computed zero-doping phonon frequency is $\omega_a^0/(2\pi c) = 1554$~cm$^{-1}$,
where $c$ is the speed of light.
The frequency variation $\Delta\omega$ with $\sigma$
is reported in Fig.~\ref{fig2}.
Calculations are done keeping the lattice-spacing constant at $a(0)$, or varying it
according to Eq.~\ref{eq2}.
In this latter case, $\Delta\omega$ is fitted by
\begin{equation}
\frac{\Delta\omega}{2\pi c} =
-2.13\sigma
-0.0360\sigma^2
-0.00329\sigma^3
-0.226|\sigma|^{3/2},
\label{eq2_5}
\end{equation}
where $\sigma$ is in 10$^{13}$~cm$^{-2}$ and $\Delta\omega/(2\pi c)$ is
in cm$^{-1}$ units.
The lattice-parameter variation is
important, since it nearly doubles the frequency shift. However,
Fig.~\ref{fig2} does not show the sudden increase of the phonon
frequency with $|\sigma|$, expected from the
displacement of the KA wavevector with the doping.
In particular, for
$\sigma=$ 3~10$^{13}$~cm$^{-2}$, the frequency variation is $\sim
-0.5\%$, which excludes a magnification effect related to the KA.

It is important to understand whether the absence of the KA is an
artifact of the adiabatic approximation, used so far.
Thus, we consider that
a phonon is not a {\it static} perturbation
but a {\it dynamic} one, oscillating at the frequency $\omega$, which
can be treated within {\it time-dependent} perturbation theory.
Using such {\it dynamic} approach in the context of DFT~\cite{allen}, the dynamical
matrix of a phonon with momentum ${\bf q}$,
projected on the phonon normal-coordinate is
\begin{eqnarray}
{\cal D}^{\epsilon_F}_{\bf q}(\omega)&=&
F^{\epsilon_F}_{\bf q}(\omega)
+\int n({\bf r}) \Delta^2V^b({\bf r}) d^3r \nonumber\\
&&-\int\Delta n^*_{\bf q}({\bf r})
K({\bf r},{\bf r'}) \Delta n_{\bf q}({\bf r'})
d^3rd^3r',
\label{eq3}
\end{eqnarray}
where $n({\bf r})$ is the charge density, 
$\Delta^2V^b$ is the second derivative of the bare (purely-ionic)
potential with respect to the phonon displacement,
$\Delta n$ is the derivative of $n$,
$K({\bf r},{\bf r'})=\delta^2E_{Hxc}[n]/
(\delta n({\bf r}) \delta n({\bf r'}))$,
$E_{Hxc}[n]$ is the Hartree and exchange-correlation functional,
and
\begin{equation}
F^{\epsilon_F}_{{\bf q}}(\omega)=
\frac{2}{N_{\bf k}}\sum_{{\bf k}nm}
\frac{|D_{({\bf k+q})m,{\bf k}n}|^2
[\tilde f_{({\bf k+q})m}-\tilde f_{{\bf k}n}]}
{\epsilon_{({\bf k+q})m}-\epsilon_{{\bf k}n}+\hbar\omega+i\delta}.
\label{eq4}
\end{equation}
Here a factor 2 accounts for spin degeneracy,
the sum is performed on $N_{\bf k}$ wavevectors,
$D_{({\bf k+q})m,{\bf k}n}=
\langle ({\bf k+q})m|\Delta V|{\bf k}n\rangle$
is the electron-phonon coupling (EPC),
$\Delta V$ is the derivative of the Kohn-Sham potential,
$|{\bf k}n\rangle$ is a Bloch eigenstate with
wavevector ${\bf k}$,
band index $n$ and energy $\epsilon_{{\bf k}n}$,
$\tilde f_{{\bf k}n}=f_T(\epsilon_{{\bf k}n}-\epsilon_F)$,
where $f_T$ is the Fermi-Dirac distribution and $\delta$
is a small real number.

Imposing $\omega=0$ and $\delta=0$
in Eq.~\ref{eq3}, one obtains the standard
adiabatic approximation~\cite{dfpt} and the phonon frequency is
$\omega^{\epsilon_F}_a=\sqrt{{\cal D}^{\epsilon_F}_{\bf q}(0)/M}$, where $M$ is the atomic mass.
In the  dynamic case, $\omega$ has to be determined
self-consistently from $\omega=\sqrt{{\cal D}^{\epsilon_F}_{\bf q}(\omega)/M}$.
However, considering dynamic and doping effects as perturbations, at the
lowest order
one can insert the adiabatic zero-doping phonon frequency $\omega_a^0$
in Eq~\ref{eq3} and obtain the real part of the dynamic frequency from
$\omega^{\epsilon_F}_d=\Re\mathrm{e}\sqrt{{\cal D}^{\epsilon_F}_{\bf q}(\omega_a^0)/M}$.

Let us consider the ${\bf q}\rightarrow {\bf 0}$ limit in Eq.~\ref{eq4}.
In the adiabatic case
\begin{eqnarray}
F^{\epsilon_F}_{\bf 0}(0)&=&
\frac{2}{N_{\bf k}} \sum_{{\bf k},n\ne m} 
\frac{|D_{{\bf k}m,{\bf k}n}|^2 [\tilde f_{{\bf k}m} - \tilde f_{{\bf k}n}] }
     {\epsilon_{{\bf k}m}-\epsilon_{{\bf k}n}} \nonumber \\
&&
-\frac{2}{N_{\bf k}} \sum_{{\bf k},n} |D_{{\bf k}n,{\bf k}n}|^2
\delta_T(\epsilon_{{\bf k}n}-\epsilon_F),
\label{eq5}
\end{eqnarray}
where $\delta_T(x)=-df_T(x)/(dx)$. In the dynamic case
\begin{eqnarray}
F^{\epsilon_F}_{\bf 0}(\omega_a^0)&=&
\frac{2}{N_{\bf k}} \sum_{{\bf k},n\ne m} 
\frac{|D_{{\bf k}m,{\bf k}n}|^2
[\tilde f_{{\bf k}m} - \tilde f_{{\bf k}n} ]}
     {\epsilon_{{\bf k}m}-\epsilon_{{\bf k}n}+\hbar\omega_a^0+i\delta}.
\label{eq6}
\end{eqnarray}
In Eq.~\ref{eq5} (adiabatic case), there are two contributions, the first
from inter-band and the second from intra-band transitions
(depending on $\delta_T$ and proportional
to the density of states at $\epsilon_F$).
On the contrary, in Eq.~\ref{eq6} (dynamic case)
only inter-band transitions contribute.

The variation of $\omega^{\epsilon_F}$ with $\epsilon_F$ is
\begin{equation}
\Delta\omega = \omega^{\epsilon_F}-\omega^0 \simeq
\frac{{\cal D}^{\epsilon_F}-{\cal D}^0}{2M\omega^0_a},
\label{eq7}
\end{equation}
where is assumed that $\Delta\omega\ll\omega^0_a$.
The presence of a Kohn anomaly is associated to a singularity in the
electron screening, which, within the present formalism, can
occur if the denominator of Eq.~\ref{eq4} approaches zero,
i.e. for electrons near the Fermi level.
Let us call $\tilde F^{\epsilon_f}(\omega)$
the part of $F^{\epsilon_F}_0(\omega)$ obtained by restricting the
{\bf k}-sum on a circle
of radius $\bar{k}$ centered on {\bf K}, with $(\beta {\bar k}-|\epsilon_F|-\hbar\omega^0_a)\gg k_{\rm B}T$.
The anomalous $\Delta\omega$ is obtained by substituting  ${\cal D}$ with $\tilde F$~\cite{nota04}
in Eq.~\ref{eq7}
\begin{eqnarray}
\Delta\omega_a&=&
\frac{\tilde F^{\epsilon_F}(0)-\tilde F^0(0)}{2M\omega^0_a} \label{eq8} \\
\Delta \omega_d &=&\Re\mathrm{e}\left[
\frac{\tilde F^{\epsilon_F}(\omega^0_a)-\tilde F^{0}(\omega^0_a)}{2M\omega^0_a}
\right]
\label{eq9}
\end{eqnarray}
in the adiabatic ($\Delta\omega_a$) and dynamic ($\Delta\omega_d$) cases.
An analytic expression for $\tilde F$ is obtained by
i) linearizing the band dispersion;
ii) writing the EPC as
$|D_{({\bf K+k})n,({\bf K+k})m}|^2=\langle D^2_{\bm \Gamma}\rangle [1\pm\cos(2\theta)]$,
where $\theta$ is the angle between the phonon-polarization and ${\bf k}$,
the sign $\pm$ depends on the transition
(see Eq.~6 and note 24 of Ref.~\cite{piscanec04}) and
$\langle D^2_{\bm \Gamma}\rangle=45.6$~(eV)$^2$/\AA$^{-2}$ from DFT
~\cite{lazzeri06};
iii) substituting
$1/N_{\bf k}\sum_{{\bf k}}$ with $2A_0/(2\pi)^2 \int d^2k$
in Eqs.~\ref{eq5}-\ref{eq6} , a factor 2 counts {\bf K}
and {\bf K'}, and ${\bf k}$ is measured from {\bf K}.

In the adiabatic case
\begin{eqnarray}
\tilde F^{\epsilon_F}(0)
&=&
\alpha\int_0^{\bar k}  kdk \left\{
\frac{f_T(\beta k-\epsilon_F)-f_T(-\beta k-\epsilon_F)}
{\beta k} \right. \nonumber\\
&&-\left.\delta_T(\beta k - \epsilon_F)
-\delta_T(-\beta k - \epsilon_F)
\right\},
\label{eq10}
\end{eqnarray}
where $\alpha=2A_0\langle D^2_{\bm \Gamma}\rangle/\pi$.
Substituting Eq.~\ref{eq10} into Eq.~\ref{eq8}
one obtains $\Delta \omega_a$. At any $T$, $\Delta \omega_a=0$.
This result is not trivial and comes from the exact cancellation of the
inter-band ($\pi$ to $\pi^*$, first line of Eq.~\ref{eq10}) and intra-band 
($\pi$ to $\pi$ and $\pi^*$ to $\pi^*$, second line
of Eq.~\ref{eq10}). For example, at $T=0$, both contributions to $\hbar \Delta \omega_a$ 
are large and equal to $\alpha' |\epsilon_F|$ and $-\alpha' |\epsilon_F|$,
respectively, 
where $\alpha'=\hbar\alpha/(2M\omega^0_a\beta^2) = 4.43~10^{-3}$
and $\alpha'/(2\pi\hbar c) =35.8$~cm$^{-1}/(eV)$.
Concluding, an adiabatic calculation of $\omega^{\epsilon_F}$
does not show any singular behavior in $\epsilon_F$ related to the Kohn anomaly,
in agreement with the state-of-the-art adiabatic DFT calculations of Fig.~\ref{fig2}.

In the dynamic case
\begin{equation}
\tilde F^{\epsilon_F}(\omega^0_a) =
\alpha\int_{-{\bar k}}^{\bar k}
\frac{f_T(\beta k-\epsilon_F)-f_T(-\beta k-\epsilon_F)}
{2\beta k + \hbar\omega^0_a + i\delta} |k|dk.
\label{eq11}
\end{equation}
Substituting Eq.~\ref{eq11} into Eq.~\ref{eq9},
for $T=0$,
\begin{equation}
\hbar\Delta\omega_d=
\alpha' |\epsilon_F| +
\frac{\alpha'\hbar\omega^0_a}{4}
\ln \left(\left|
\frac{|\epsilon_F|-\frac{\hbar\omega^0_a}{2}}{|\epsilon_F|+\frac{\hbar\omega^0_a}{2}}
\right|\right).
\end{equation}
In this case, the situation
is very different since the large inter-band contribution is not canceled
by an intra-band term. In particular, there are two logarithmic divergences for
$\epsilon_F=\pm \hbar\omega^0_a/2$ and for
$|\epsilon_F|\gg\hbar\omega^0_a/2$ the frequency increases as
$\alpha'|\epsilon_F|$.

\begin{figure} [h]
\centerline{\includegraphics[width=65mm]{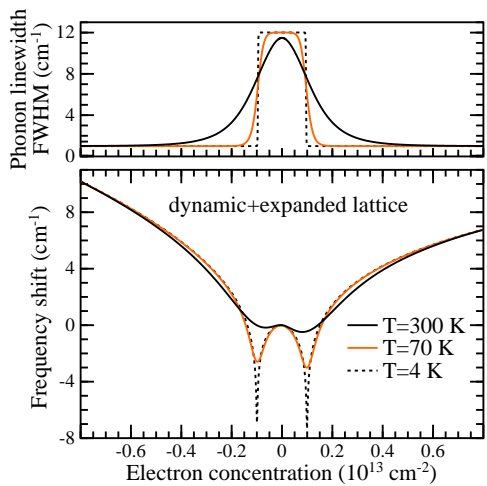}}
\caption{(Color online)
Linewidth and dynamic frequency of the E$_{2g}$ ${\bm \Gamma}$ mode
(Raman $G$ band).
See the caption of Fig.~\protect\ref{fig1}.
}
\label{fig3}
\end{figure}

$\Delta\omega_d$ computed in this way takes into account transitions
between states close to the Fermi level. However, the frequency is
also affected by the variation of the lattice-spacing, by the
transitions involving a state far from $\epsilon_F$ and by the second
and third terms in Eq.~\ref{eq3}.  All these contributions are
accurately described by our adiabatic DFT calculations.  Therefore, to
compare with experiments, we add $\Delta\omega_d$ to the adiabatic DFT
frequency shift of Eq.~\ref{eq2_5}.  The results are shown in
Fig.~\ref{fig2} for $T=300~K$, and in Fig.~\ref{fig3} as a function of
$T$ for a smaller $\sigma$ range.  Even at room temperature, the
non-adiabatic Kohn anomaly magnifies the effect of the doping and for
a valence-charge variation of -0.2\% (+0.2\%), the frequency varies by
+1.5\% (+0.7\%).  $\Delta\omega$ is asymmetric with respect to
$\epsilon_F$ and has a maximum for $\sigma\sim+3.5~10^{13}$~cm$^{-2}$.
Since $\Delta\omega_d$ is an even function of $\epsilon_F$, this lack
of electron-hole symmetry is entirely due to the adiabatic DFT
contribution.  The $\epsilon_F=\pm\hbar\omega^0_a/2$ logarithmic anomalies
are visible at $T=4$~and$~70$~K.
The presence of a logarithmic KA in this two-dimensional system is
quite remarkable since such divergences are typical of
one-dimensional systems.  They are present in graphene because of its
particular massless Dirac-like electron band dispersion.

Finally, the Raman $G$-band has a finite homogeneous linewidth due to the
decay of the phonon into electron-hole pairs.
Such EPC broadening
can be obtained either from the imaginary part of the TDPT dynamical
matrix (Eq.~\ref{eq11}) or, equivalently, from the Fermi golden rule~\cite{lazzeri06}:
\begin{equation}
\gamma = \frac{\pi}{2}\frac{\omega_a^0}{2\pi c}\alpha'
\left[f_T\left(-\frac{\hbar\omega_a^0}{2}-\epsilon_F\right)-
f_T\left(\frac{\hbar\omega_a^0}{2}-\epsilon_F\right) \right]
\end{equation}
where $\gamma$ is the full-width half-maximum (FWHM) in cm$^{-1}$.
At $T=0$ and $\epsilon_F=0$, one recovers the result of Ref.~\cite{lazzeri06},
$\gamma=11.0$~cm$^{-1}$.
The phonon-phonon scattering contribution to the FWHM
is smaller ($\sim$1~cm$^{-1}$~\cite{bonini}) and independent of $\epsilon_F$.
The total homogeneous FWHM is reported in Fig.~\ref{fig3}.
The FWHM displays a strong doping dependence; it suddenly drops for $|\sigma|\sim 0.1~10^{13}$~cm$^{-2}$
($|\epsilon_F|\sim 0.1$~eV).
Indeed, because of the energy and momentum conservation, a ${\bm \Gamma}$
phonon decays into one electron (hole) with energy $\hbar\omega_a^0/2$ above (below)
the level crossing.
At $T=0K$ such process is compatible with the Pauli exclusion-principle only
if $|\epsilon_F|<\hbar\omega_a^0/2$.

Concluding, a Kohn anomaly dictates the dependence of the highest optical-phonon
on the wavevector ${\bf q}$, in undoped graphene ~\cite{piscanec04}.
Here, we studied the impact of such anomaly on the ${\bf q=0}$
phonon, as a function of the charge-doping $\sigma$.
We computed, from first-principles, the phonon frequency and linewidth of the 
E$_{2g}$, ${\bm \Gamma}$ phonon (Raman $G$ band)
in the $\sigma$-range reached by recent FET experiments.
Calculations are done using i) the customary {\it adiabatic} Born-Oppenheimer approximation
and ii) time-dependent perturbation theory to explore {\it dynamic} effects beyond
this approximation. The two approaches provide very different results.
The {\it adiabatic} phonon frequency displays a smooth dependence on $\sigma$ and
it is not affected by the Kohn anomaly.
On the contrary, when {\it dynamic} effects are included,
the phonon frequency and lifetime display a strong  dependence on $\sigma$,
due to the Kohn anomaly.
The variation of the Raman $G$-band with the doping in a graphene-FET
has been recently measured by two groups~\cite{kim07,ferrari07}.
Both experiments are well described by our {\it dynamic} calculation but not by the 
more approximate {\it adiabatic} one.
We remark that the adiabatic Born-Oppenheimer approximation is
considered valid in most materials and is commonly used for phonon
calculations.  Here, we have shown that doped graphene is a spectacular
example where this approximation miserably fails.

We aknowledge useful discussions with A.M. Saitta, A.C. Ferrari and S. Piscanec.
Calculation were done at IDRIS (Orsay, France), project n$^o$ 061202.


\begin{thebibliography}{99}
\bibitem{nota1}
Consider a graphene sample of $N$ unit-cells, which is in contact
with an electrode of area $A^e$. If $N>A^e/A$, $A$ is obtained by
minimizing the total energy of the sample
$E(\sigma,A)A^e/A+(N-A^e/A)E(0,A_0)$. This
is equivalent to minimizing $F(\sigma,A)$.
\bibitem{nota2}
By a simple electrostatic model, the dependence of the cell-energy $E$ on 
$L$ is
$E(L)/A=\pi/6\sigma^2L+\alpha_0+\alpha_1/L+\alpha_2/L^2+{\cal O}(1/L^3)$.
$E$ is computed for $L=7.5, 15, 22.5, 30, 37.5$~\AA~and the parameters $\alpha_i$,
are fitted to obtain 
the limit $L\rightarrow\infty$.
\bibitem{nota04}
The dependence of $\Delta n$ on $\omega_{\bf q}$ and $\epsilon_F$
is neglected, because the functional of Eq.~\ref{eq3} is stationary with
respect to $\Delta n$.
Notice that there are other, equivalent expressions for the dynamical matrix 
which are not stationary in $\omega_{\bf q}$, for which this approximation
is not justified.
\end{thebibliography}
\end{document}